\documentclass[floatfix,aps,prd,showpacs,amsmath,nofootinbib,preprintnumbers,
twocolumn]
{revtex4}
\usepackage{graphicx}
\usepackage{colordvi}

    \def\be{\begin{equation}}
    \def\ee{\end{equation}}
    \def\ba{\begin{eqnarray}}
    \def\ea{\end{eqnarray}}

\begin{document}

\title{Generation of fluctuations during inflation: comparison of
stochastic and field-theoretic approaches}

\author{F. Finelli$\,^{1,2,3}$, G. Marozzi$\,^{4,3}$, A. A. 
Starobinsky$\,^{5}$, 
G. P. Vacca$\,^{4,3}$ and G. Venturi$\,^{4,3}$}

\affiliation{$^{1}$ INAF/IASF Bologna,
Istituto di Astrofisica Spaziale e Fisica
Cosmica di Bologna \\
via Gobetti 101, I-40129 Bologna - Italy}
\affiliation{$^{2}$ INAF/OAB, Osservatorio Astronomico di Bologna,
via Ranzani 1, I-40127 Bologna -
Italy}
\affiliation{$^{3}$ INFN, Sezione di Bologna,
Via Irnerio 46, I-40126 Bologna, Italy}
\affiliation{$^{4}$ Dipartimento di Fisica, Universit\`a degli Studi di 
Bologna, via Irnerio, 46 -- I-40126 Bologna -- Italy}
\affiliation{$^{5}$ Landau Institute for Theoretical Physics, Moscow, 119334, 
Russia}

\begin{abstract}
We prove that the stochastic and standard field-theoretical approaches 
produce exactly the same results for the amount of light massive scalar 
field fluctuations generated during inflation in the leading order of the 
slow-roll approximation. This is true both in the case for which this field 
is a test one and inflation is driven by another field, and the case for which
the field plays the role of inflaton itself. In the latter case, in order 
to calculate the mean square of the gauge-invariant metric fluctuations, 
the logarithm of the scale factor $a$ has be used as the time variable in the 
Fokker-Planck equation in the stochastic approach. The implications of particle 
production during inflation for the second stage of inflation and for the 
moduli problem are also discussed. The case of a massless self-interacting 
test scalar field in de Sitter background with a zero initial 
renormalized mean square is also considered in order to show how the 
stochastic approach can easily produce results corresponding to diagrams 
with an arbitrary number of scalar field loops in the field-theoretical 
approach (explicit results up to 4 loops included are presented).
\end{abstract}
\maketitle

\section{Introduction}
It has been known for a long time that light minimally coupled scalar
fields typically have anomalously high vacuum expectation values 
(VEV) for even powers of fields in the de Sitter background. In 
particular, the mean square of a free, massive and minimally coupled to 
gravity scalar field in the equilibrium, de Sitter invariant quantum 
state (the Bunch-Davies vacuum) is \cite{bd_book} 
\be
\langle \phi^2 \rangle \simeq \frac{3 H^4}{8 \pi^2 m^2} \gg H^2
\label{bd}
\ee
if $m^2\ll H^2$. Here $H\equiv \dot a/a$ is the Hubble parameter and 
$a(t)$ is the scale factor of a Friedmann-Robertson-Walker (FRW) 
cosmological model. For the particular case $m=0$ and 
$\langle \phi^2 \rangle=0$ (or small) at the beginning of the de Sitter 
stage (set at $t=0$ here), we have 
\be 
\langle \phi^2 \rangle \simeq \frac{H^3 t}{4 \pi^2}
\label{af}
\ee
which grows without bound \cite{L82,S82,VF82}. This anomalous growth is 
an essentially infrared effect, it occurs due to field Fourier modes with
wavelengths far exceeding the de Sitter event horizon. Thus, it is not
a consequence of the Gibbons-Hawking effective temperature $T=H/2\pi$ 
\cite{GH77} experienced by a point observer inside her/his de Sitter 
horizon. In particular, in contrast with temperature-like effects which 
should be universal for fields of all spins, this effect occurs for 
minimally coupled light scalar fields and gravitons only (but not for, 
e.g., photons or light fermions).

Because of continuous creation of infrared modes and growth of their 
occupation number, the quantum scalar field can be split into a long 
wave (coarse grained) component and a short-wave (perturbative) one.
Then it can be proved that the former component effectively becomes
quasi-classical, though random (i.e. all non-commutative parts of it
may be neglected), and it experiences a random walk described by 
the stochastic inflation approach. In some specific case but beyond 
one-loop approximation, this approach was used already in \cite{S82}; 
see \cite{S86} for the rigorous derivation in the generic case when this 
scalar field is a slow-rolling inflaton itself and for a number of 
analytic non-perturbative results.\footnote {Description of the
growth (\ref{af}) in the free massless case in terms of the Fokker-Planck 
equation was first considered in \cite{V83}. The stochastic approach was
applied to the description of eternal inflation in \cite{L86}.} The
case of a scalar field with the quartic self-interaction in the exact
de Sitter space-time was first studied using this approach in \cite{SY}, 
and a number of non-perturbative results beyond any finite number of
loops in quantum field theory (QFT) perturbative expansion were obtained 
there.

However, it should be emphasized that just because of the non-perturbative
nature of the stochastic approach to inflation, it is based on a number 
of heuristic approximations. Therefore, it is very important to check,
whenever possible, results obtained by its application using the standard
perturbative QFT in curved space-time. Also, an inflationary space-time
is not the exactly de Sitter one, $\dot H\not=0$, that can often lead
to a drastic change in conclusions. That is why our purpose here is to  
consider new applications of this approach to inflationary space-times 
with $\dot H \ne 0$, though with $|\dot H|\ll H^2$ (as required by 
observational data on a slope of the primordial power spectrum of 
scalar perturbations in the Universe), and to compare results obtained 
in this way with those directly following from perturbative QFT in curved 
space-time. Our paper contains several novel results referring both to 
inflationary space-times and to the exact de Sitter space-time beyond
one-loop approximation. Also, our results shed new light on the much
discussed problem of the choice of an independent time variable in
the Langevin and Fokker-Planck equations in the stochastic approach.
Finally, we show that in most cases the mean square of a scalar field 
at last stages of inflation is very different from the instantaneous
Bunch-Davies value (\ref{bd}) -- even ``eternal'' inflation (which does
occur in the inflationary models we consider for a sufficiently large
initial value of an inflaton field) is not eternal enough for a light 
scalar field to reach equilibrium. 

The paper is organized as follows. In Sec. II the growth of fluctuations of
a light test field during inflation driven by a massive inflaton is
investigated. We prove that the results obtained in \cite{FMVV_I,FMVV_II} 
using QFT with adiabatic regularization in curved space-time can be 
obtained from a general diffusion equation with a noise term which has 
the same form as that in the Langevin equation for an inflaton field 
in the stochastic inflation approach \cite{S86}. In Sec. III more details 
are presented on how the mean square of a test scalar field with an 
arbitrary (though light, $m \ll H$) mass is obtained using adiabatic 
regularization in curved space-time with a slowly changing curvature. 
Generation of light scalar field fluctuations in inflationary models can 
be so strong as to dominate finally the classical energy density of an 
inflaton and to drive a second stage of inflation. As an example, in 
Sec. IV we derive the conditions under which such a second stage can occur 
for the $m^2 \phi^2$ inflationary model, with or without a break between 
the two stages of inflation. In Sec. V we discuss the impact of this 
inflationary particle production on the moduli problem. We show that 
quantum moduli problems are worse than the classical one and we 
improve on previous investigations \cite{GLV,FKL}. In Sec. VI we discuss 
the diffusion (Langevin) equation for inflaton fluctuations in 
case of a generic chaotic power-law potential $V(\phi) \propto \phi^n$. 
Using the results obtained by QFT methods, we show that if one is 
interested in metric fluctuations, this diffusion equation should be 
formulated in terms of the independent time variable $\ln a$ which 
is directly related to the number of e-folds during inflation $N$
(this was among the possibilities envisioned in \cite{S86}). In Sec. VII 
it is shown that the stochastic method can also be used beyond the 
one-loop approximations for a self-interacting scalar field in the exact 
de Sitter space-time, again leading to the same results as obtained by 
QFT methods. We conclude in Sec. VIII and discuss the perturbative 
expansion of the diffusion equation in the Appendix.

\section{Growth of light test field fluctuations 
during massive inflation in stochastic approach}
Let us first consider the production of particles and fluctuations for a test
quantum field $\chi$ with a small mass $m\ll H$. Thus, we neglect the 
$\chi$ energy density and pressure in the background FRW equations. Also, 
it is assumed that there is no Bose-condensate of $\chi$, $<\chi>=0$.

As an example, in this section we limit ourselves to the simplest case of 
inflaton with the quadratic potential. Some of QFT results for this model 
have already been obtained in \cite{FMVV_I,FMVV_II}. The Friedmann 
equation for the massive inflaton is:
\be
H^2 = \frac{\rho_\phi}{3 M_{\rm pl}^{2}} = 
\frac{1}{3 M_{\rm pl}^{2}} \left[ \frac{\dot\phi^2}{2} + m^2
\frac{\phi^2}{2} \right] \simeq \frac{m^2}{M_{\rm pl}^{2}}
\frac{\phi^2}{6}
\ee
where $8 \pi G = M_{\rm pl}^{-2}$. In the slow-roll approximation,
the inflationary trajectory is well approximated by
\begin{eqnarray} \label{ansatz2}
H&=&H (t) \simeq H_0 - \frac{m^2}{3} (t-t_0)\,,
\\
\phi (t) &\simeq& \phi_0 - \sqrt{\frac{2}{3}} m M_{\rm pl} (t - t_0)\,, \\
a(t) 
&\simeq& a_0 \exp \left[ \frac{3}{2 m^2} \left( H_0^2 - H^2 \right) \right]\,,
\end{eqnarray}
where the subscript $0$ denotes the beginning of inflation (these
expressions were first presented in \cite{S78} in the context of a closed 
bouncing FRW universe). In the rest of the paper, we set $a_0=1$ for 
simplicity. The equation for inhomogeneous Fourier modes of $\chi$ is:
\be
\ddot \chi_k + 3 H \dot \chi_k + \left[ \frac{k^2}{a^2} + m_\chi^2 \right]
\chi_k= 0 \,.
\label{eqmodes}
\ee
and we denote $m_\chi^2=\alpha \,m^2$. Let us first discuss the case  
$m_\chi = m$, or $\alpha=1$, in order to see the relation between
renormalized QFT results and the stochastic approach. This case was 
studied in \cite{FMVV_I} where inflaton fluctuations in a rigid space-time 
were considered. For the value of $\langle \chi^2 \rangle$ renormalized 
by adiabatic subtraction, one obtains in the leading order:
\be
\langle \chi^2 \rangle_{\rm REN} \simeq \frac{H^2}{4 \pi^2} \log a
\,,
\label{leading_same}
\ee

This result agrees with the solution of the differential equation for 
$\langle\chi^2 \rangle_{\rm REN}$:
\be
\frac{d \langle \chi^2 \rangle_{\rm REN}}{d t} 
+ \frac{2 m^2}{3 H (t)} \langle \chi^2 \rangle_{\rm REN} =
\frac{H^3 (t)}{4 \pi^2} \,,
\label{stochastic}
\ee
(see e.g. Eq. (10) of paper \cite{S82}) with $H$ not constant in
time, but having the time evolution given by Eq. (\ref{ansatz2}).
In the scope of the stochastic approach, the right-hand side of
Eq. (\ref{ansatz2}) arises from a noise term in the Langevin equation
for the large-scale part of $\chi$, see Eq. (\ref{phi_general_stoch}) 
below, coarse-grained over the physical 3D-volume $\sim(\epsilon H)^{-3}$
which slowly expands during inflation. Here $\epsilon \ll 1$ is an 
auxiliary parameter which enters into the definition of the 
coarse-graining scale $k_{\rm cg}=\epsilon a H$ in momentum space. 
Since we are interested in the leading order of the slow-roll 
approximation, terms of order $\dot H/H^2$ coming either from taking an 
explicit time derivative of $H$ or from solutions of the wave equation 
(\ref{eqmodes}) at the course-graining scale may be neglected. With the 
same accuracy, one may alternatively take the physical course-graining 
scale $k_{\rm cg}/a$ to be exactly constant during inflation and equal 
to $\epsilon H(t_f)$ where $t_f$ denotes the end of inflation.

The general solution to Eq. (\ref{stochastic}) in the
background (\ref{ansatz2}) is indeed:
\ba
\langle \chi^2 \rangle_{\rm REN} &=& C H^2 + \frac{H^2}{4 \pi^2} \log a 
\nonumber \\ &=& C H^2 + \frac{3 H^2}{8 m^2 \pi^2} \left( H_0^2 - H^2 
\right)\,,
\ea
where $C$ is an integration constant and we have used Eq. (6) in the 
second expression. For a long stage of inflation, when
$H(t) << H_0$, we have \cite{FMVV_I}:
\ba
\langle \chi^2 \rangle_{\rm REN} &\simeq& C H^2 + \frac{3 H^2}{8 m^2
\pi^2} H_0^2 \nonumber \\
&=& \langle \chi^2 (t_0) \rangle_{\rm REN}
\frac{H^2}{H_0^2} + \frac{3 H^2}{8 m^2
\pi^2} H_0^2\,.
\ea
The regime $H(t) << H_0$ occurs only when the number of e-folds is
$N \approx N_0 = 3 H_0^2/(2 m^2)$.

The stochastic equation (\ref{stochastic}) can be easily extended to
a generic $m_\chi$ not coinciding with the inflaton mass $m$:
\be
\frac{d \langle \chi^2 \rangle_{\rm REN}}{d t} 
+ \frac{2 m_\chi^2}{3 H (t)} \langle \chi^2 \rangle_{\rm REN} =
\frac{H^3 (t)}{4 \pi^2} \,.
\label{stochastic_generic}
\ee
Its solution is:
\begin{eqnarray}
\!\!\!\!\!\!\!\!\!\!\!\langle \chi^2 \rangle_{\rm REN} \!\!&=& \!\!C_1 H^{2
  \alpha} + \frac{3 H^4}{8 \pi^2 m^2 (\alpha-2)} \nonumber \\
&=& \!\!C_2 H^{2 \alpha} + \frac{3 H^{2 \alpha}}{8
\pi^2 m^2 (2-\alpha)} ( H_0^{4-2 \alpha} \!- \!H^{4-2 \alpha} )
\label{soldiff}
\end{eqnarray}
where $C_1 \,, C_2$ are integration constants.
For the particular marginal value $\alpha=2$, 
Eq. (\ref{soldiff}) should be replaced by:
\be
\langle \chi^2 \rangle_{\rm REN} = C_2 H^{4} - \frac{3 H^4}{4
\pi^2 m^2} \log \left(\frac{H}{H_0}\right) \,.
\label{alfaeq2}
\ee


It is important to note that $\langle \chi^2 \rangle_{\rm REN}$ is 
different from the instantaneous Bunch-Davies equilibrium
value $3 H^4 / (8 \pi^2 m_\chi^2)$ in Eq. (\ref{bd}) for any $\alpha$. 
Since $H (t)$ decreases with time, it is only for $\alpha >> 2$ that 
$\langle \chi^2 \rangle_{\rm REN}$ may approach (but not become exactly 
equal to) the instantaneous Bunch-Davies vacuum value at the last stage of 
inflation.

When $H (t) << H_0$ and $\alpha<2$,
\be
\langle \chi^2 \rangle_{\rm REN} \simeq \langle \chi^2 (t_0) \rangle_{\rm
REN} \frac{H^{2 \alpha}}{H_0^{2 \alpha}}
+ \frac{3 H^{2 \alpha} H_0^{4 - 2 \alpha}}{8 m^2
\pi^2 (2-\alpha)} \,.
\label{xi2limit}
\ee
In the limit $\alpha << 2$ (but $\alpha \ne 0$), $\chi$ renormalized
fluctuations - and therefore the $\chi$ energy density - depend 
non-trivially on the duration of inflation \cite{FMVV_I,FMVV_II}. 
Note that only for $\alpha=1$ (i.e. $m_\chi=m$) and at the beginning of 
inflation, $< \chi^2 > \sim H_0^3 t$ as occurs in the exact de Sitter 
space-time.

This shows the occurrence of a new characteristic scale in the problem, 
namely $\sqrt{-\dot H}$. In contrast to the de Sitter space-time, where 
$H$ is the only scale present, in inflation we have also $\sqrt{-\dot H}$, 
i.e. $\frac{m}{\sqrt{3}}$ in the model under consideration. Fields which 
are light compared to the Hubble parameter, are subsequently divided in 
two classes: the ones with $m_\chi > \sqrt{-3 \dot H}$ and the ones with 
$m_\chi < \sqrt{-3 \dot H}$. Their corresponding particle production 
rates look different.

A homogeneous solution of Eq. (\ref{stochastic_generic}) given by the
term containing $C_1$ in  Eq. (\ref{soldiff}) is also the slow-roll 
solution for $\chi_{\rm cl}^2$, with $\chi_{\rm cl}$ being the classical 
homogeneous field $\chi(t)$. In the case $\alpha<2$, this part of the
general solution becomes negligible with respect to generated quantum 
fluctuations as inflation develops if
$\langle \chi^2 (t_0) \rangle_{\rm REN} << 3 H_0^4/\left(8 \pi^2 m^2
(2-\alpha) \right)$ (see Eq. (\ref{xi2limit})). In particular, if zero 
initially, $\langle \chi^2 \rangle_{\rm REN}$ becomes larger than the 
instantaneous Bunch-Davies equilibrium value when
\be
\frac{H}{H_0} < \left( \frac{\alpha}{2 -\alpha} \right)^\frac
{1}{2(2-\alpha)}\,.
\ee
The energy density of $\chi$ becomes comparable to that of the inflaton 
and cannot be neglected further on when
\be
\frac{\rho_\chi}{\rho_\phi} \ge
\frac{m_\chi^2 \langle \chi^2 \rangle_{\rm REN}}{6 H^2 M_{\rm pl}^2} \sim 1 \,.
\label{chi_energy}
\ee
This never happens for $\alpha\ge1$, but for $\alpha<1$ back-reaction 
effects become important when
\be
H(t) \le H(t_{\rm \star})= H_0 \left( \frac{H_0 \sqrt{\alpha}}{4 \pi 
\sqrt{2-\alpha}
M_{\rm pl}}
\right)^\frac{1}{1-\alpha} \,.
\label{tstar}
\ee
Note that the dominant term in $\rho_{\chi}$ is $m^2 \langle \chi^2 
\rangle_{\rm REN}$, the $\chi$ kinetic and gradient energy are sub-leading.  
Also, we always consider $H_0<\gamma M_{\rm pl}$ with $\gamma={\cal O}(1)$.

\section{Renormalization approach}
We now wish to present the renormalized $\langle \chi^2 \rangle_{\rm REN}$
for a generic mass $m_\chi$ obtained using dimensional regularization and 
adiabatic subtraction. Such a calculation goes beyond \cite{FMVV_I}, 
which addressed the first case in the previous section ($\alpha=1$) only.

Let us consider solutions of the wave equation Eq. (\ref{eqmodes}) which 
give time-dependent coefficients in the expansion of the quantum field 
$\chi$ in terms of the Fock annihilation and creation operators for each 
Fourier mode ${\bf k},\, k=|{\bf k}|$. We use 
a set of approximate solutions of Eq. (\ref{eqmodes}), chosen so as to 
separate the ultraviolet (UV henceforth) and infrared (IR henceforth) 
domains for $k> {\bar \epsilon} a H$ and $k< {\bar \epsilon} a H$ 
respectively (here ${\bar \epsilon}<1$ need not coincide with the 
parameter $\epsilon$ in the definition of the coarse graining scale 
introduced above).

The UV solution whose contribution will mostly be removed by the 
adiabatic regularization can be written in the generic form
\be
\chi^{\rm UV}_k=\frac{1}{a^{3/2}} \left( \frac{\pi \lambda}{4 H}
\right)^{1/2}
H^{(1)}_\nu\left( \frac{\lambda k}{a H}\right)
\ee
for suitable functions $\nu$ and $\lambda$ (whose values are normally 
close to $3/2$ and $1$ respectively, which are the values in the lowest 
order of the slow-roll approximation).

The most important contribution is the far IR one. Such a solution can 
be computed in the same approximation as used in our previous work
\cite{FMVV_II}, where a match at the moment $t_k$ when 
$k= {\bar \epsilon} a(t_k) H(t_k)$ with the UV solution is imposed, 
together with the requirement of being a solution of (\ref{eqmodes}) in 
the deep IR limit $k \to 0$. In the leading order, we choose the form
\be
\chi^{IR}_k=\frac{1}{a^{3/2}(t)}
\left( \frac{\pi \lambda}{4 H(t)} \right)^{1/2}
\left( \frac{H(t_k)}{H(t)} \right)^x
H^{(1)}_\frac{3}{2}\left( \frac{\lambda k}{a(t) H(t)}\right)
\label{IRapprox}
\ee
where the parameter $x$ can be fixed using the second constraint.
Note for the later use that at the moment $t_k$ one has
\be
H(t_k) = H_0 \sqrt{1+2\frac{\dot{H}}{H_0^2} \log{\frac{k}{\bar{\epsilon} 
H(t_k)}} }
\simeq H_0 \sqrt{1+2\frac{\dot{H}}{H_0^2} \log{\frac{k}{\bar{\epsilon} 
H_0}} } \,.
\label{deffHtk}
\ee
Substituting Eq. (\ref{IRapprox}) into Eq. (\ref{eqmodes}) 
and considering the deep IR region, one finds that $x=1-\alpha$.

In order to compute the renormalized value of $\langle \chi^2 \rangle$
at the same scale as is used in the stochastic approach, we note
that the contribution of IR modes should be taken up to 
$k=\epsilon a(t) H(t)$ (the UV ones cancel due to adiabatic subtraction). 
Then it is natural to define $t_k$ by using $\bar{\epsilon}=\epsilon$ in
Eq. (\ref{deffHtk}) in order to avoid terms like 
$\log{\bar{\epsilon}/\epsilon}$. We shall therefore make this choice.

As mentioned above, the leading contribution to the renormalized value 
of $\langle \chi^2 \rangle$ is given by the integration over IR modes. 
In the leading order of the slow-roll approximation, one obtains the 
following result:
\ba
\!\!\!\!\!\!\!\!\!\!\!\!\!\!\langle \chi^2 \rangle_{REN} 
&\simeq& \langle \chi^2 \rangle_{IR} \nonumber \\
&\simeq& \frac{H^{2\alpha}(t)}{4\pi^2}
\int_{\log{l}}^{\log{\epsilon \,a(t) H(t)}} 
H^{2-2\alpha}(t_k) \, d \log{k}  ,
\label{ren_value}
\ea
with $l={\epsilon} H_0$ and where the expansion of the Bessel function 
for a small argument is performed. Thus, adiabatic subtraction cancels 
the UV contribution more and more accurately as inflation develops. This 
defines a dynamical cutoff, and the IR part below this cut-off soon 
becomes dominant. Taking the time derivative of Eq. (\ref{ren_value}), 
one obtains:
\be
\frac{d}{d t} \langle 0| \chi^2 |0 \rangle_{REN} = \frac{H^3 (t)}{4\pi^2}
+ 2\alpha \frac{\dot{H}}{H} \langle 0| \chi^2 |0 \rangle_{REN} \, .
\label{diff_eq_alpha}
\ee
This coincides with Eq. (\ref{stochastic_generic}) for a generic mass 
$m_\chi^2=\alpha \, m^2$ since the direct computation of the integral in 
Eq. (\ref{ren_value}) in the slow-roll approximation leads to the second 
expression in  Eq. (\ref{soldiff}) with $C_2=0$. Therefore, the two
methods of doing this calculation agree.

\section{Consequences of Particle Production: second stage of 
inflation}
Let us discuss in more detail how the back-reaction of $\chi$
may become important. If $\alpha \ge 1$, then the back-reaction of $\chi$ 
will never be important during inflation (and afterwards, too, if the 
inflaton does not decay faster than $\chi$ during (p)reheating). So, we
consider $\alpha < 1$ in the following.

Let us introduce a new quantity $\beta = m^2 M_{\rm pl}^2 / H_0^4$. 
Back-reaction of $\chi$ becomes important before or after the end of
inflation driven by the inflaton $\phi$ if $H(t_*)$ in Eq. (\ref{tstar}) 
is greater or smaller than $m$ (which is the scale of the end of inflation 
in the $m^2 \phi^2$ inflationary model) respectively. Therefore, $\chi$ 
fluctuations become important before the end of inflation if
\be
\left( \frac{H_0}{m} \right)^{2 (1-\alpha)} \frac{H_0^2}{16 \pi^2 M_{\rm
pl}^2} \frac{\alpha}{2-\alpha} > 1 \,.
\ee
Thus, for $\alpha << 1$ there is a second stage of inflation without a
break if 
\be
\frac{1}{\beta} = \frac{H_0^4}{m^2 M_{\rm pl}^2} > \frac{32 \pi^2}{\alpha} \,,
\ee
and with a break if the $<$ holds in the equation above. A second stage 
of inflation without a break starts in the presence of a light field with 
mass $m_\chi$ in the range
\be
1 >> \alpha = \frac{m^2_\chi}{m^2} > 72 \pi^2
\frac{M_{\rm pl}^2}{m^2 N_0^2}
\rightarrow m^2_\chi > 72 \pi^2
\frac{M_{\rm pl}^2}{N_0^2}
\,,
\ee
which is a very broad range for large enough $N_0$. A second stage of 
inflation with a break would occur for smaller $m_\chi$. In such a case the 
slow-roll of the inflaton field practically ends, but inflation would be 
sustained by quantum fluctuations of $\chi$ corresponding to a much lower 
value for $|\dot{H}|<<m^2/3$ (see also \cite{KLS85,GMS91,FVV} for examples 
of inflation as a whole, or its second stage driven by quantum fluctuations
of a scalar field). Note that a second stage of inflation can occur even 
if curvature during the first stage of inflation is low, $H_0 << M_{\rm pl}$,
usually after a break in this case. Thus, a second stage of inflation seems 
to be very probable if the inflaton is not the lightest field around. Also, 
even without the second stage, $\chi$ may play a role of a curvaton 
\cite{LM,MT01,ES02,LW}.

\section{Consequences of Particle Production: Impact on the 
Moduli Problem}
On considering $\chi$ as a modulus, the above calculations allow a better 
estimate of the moduli abundance
produced by the expansion during inflation. The previous section shows 
that the moduli can even drive a second inflationary stage if $\alpha < 1$ 
and the energy stored will dominate before the end of inflation 
(see Eq. (\ref{tstar})). We focus here on the case for which 
$\chi$ does not dominate during inflation. 

In order to estimate the
ratio $n_\chi/s$ (where $n_\chi=\rho_\chi/m_\chi$ and $s$
are the number of $\chi$ particles and the entropy of produced particles,
respectively), which is required to be less than $10^{-12}$, we proceed as
in \cite{FKL}. We assume the immediate
thermalization for the inflaton and $\chi$ after the accelerated expansion:
\ba
\rho_\phi &=& 3 H^2 M_{\rm pl}^2 \simeq  \frac{\pi^2 g T^4}{30}\,, 
\nonumber \\
s &=&  \frac{2 \pi^2 g T^3}{45} \,,
\ea
where $T$ is the reheating temperature and $g$ is the number of 
species (for instance $g=106.75$ for the
standard model \cite{KT}). 

We first consider $\alpha >>2$ and obtain:
\be
\frac{n_\chi}{s} \simeq \frac{5 g}{96 \times 10^2} 
\frac{T^5}{m_\chi M_{\rm pl}^4} \,.
\ee
Because of strong positive dependence on the reheating temperature, 
the requirement $n_\chi/s < 10^{-12}$ can easily be satisfied. 

We now consider $\alpha < 2$, the case for which particle production 
is very different from the one extrapolated from the exact de Sitter 
space-time. The energy density stored in $\chi$ for $H << H_0$ is:
\be
\rho_\chi \simeq m_\chi^2 \frac{\langle \chi^2 \rangle_{\rm REN}}{2}
\simeq \alpha \frac{3 H^{2 \alpha} H_0^{4 - 2 \alpha}}{16 \pi^2
(2-\alpha)} \,,
\ee
and we obtain:
\ba
\frac{n_\chi}{s} &\simeq& \frac{3}{64 \pi^2}
\frac{\alpha T}{m_\chi} \frac{H^{2\alpha-2}}{2-\alpha}
\frac{H_0^{4-2\alpha}}{M_{\rm
pl}^2} \nonumber \\
& \propto &  T^{4 \alpha-3}
\,.
\label{ratio}
\ea
This equation replaces Eq. (4.16) of \cite{FKL}.
\footnote{ Eq. (4.16) of \cite{FKL} agrees with Eq. (\ref{ratio})
for $\alpha =1$ only, nevertheless, it is used for any $\alpha$.  
Also, the growth law (\ref{af}) is used there which is 
valid for $\alpha=1$ only as pointed above.}
Note that the dependence on
$T$ is rather peculiar: for $\alpha > 3/4$, it grows with $T$, while for
$\alpha < 3/4$, it decreases with $T$. For $\alpha \sim 0$, the ratio is
\be
\frac{n_\chi}{s} \sim \alpha \frac{270}{128 \pi^4} \frac{H_0^4}{m_\chi
\, g \,T^3} \,.
\ee
Therefore the problem of light moduli cannot be solved by lowering 
the reheating temperature in the $m^2 \phi^2$ model and is much worse 
than expected in \cite{FKL}.

\section{Inflaton Fluctuations and Stochastic Approach for Gauge 
Invariant Fluctuations}
An inflaton has an effective mass which is much smaller than the Hubble 
parameter during inflation. Therefore, the above approach should hold for 
inflaton fluctuations, too. However, such a case differs from the
previous one, since scalar metric fluctuations should be taken into
account in addition to field (inflaton) ones. In other words, the quantity
which is quantized is a linear combination of field and metric 
fluctuations. As in \cite{FMVV_II}, we choose the gauge in which inflaton 
fluctuations coincide with the gauge-invariant Mukhanov variable which is 
canonically quantized \cite{M85}. Then our results will be valid in any 
gauge, if a gauge invariant variable is considered. 

The time evolution of the renormalized inflaton fluctuations for the
$m^2\phi^2$ inflationary model has been already obtained in 
\cite{FMVV_II,FMVV_IV} using a perturbative QFT analysis of the Einstein 
equations. To the lowest order in the slow-roll and in long-wavelength 
approximations, equations for fluctuations in the first \cite{FMVV_II} and 
second \cite{FMVV_IV} order are: 
\be
3 H \dot{\delta \phi^{(1)}} +
\left(m^2 + 6\dot{H}\right) \delta \phi^{(1)} \approx 0 \,,
\label{first_order}
\ee
\be
3 H \dot{\delta \phi^{(2)}} +
\left(m^2 + 6\dot{H}\right) \delta \phi^{(2)}
  \approx \frac{\dot{\phi}}{2H} \frac{m^2}{M_{pl}^2} \left(\delta
    \phi^{(1)}\right)^2 \,.
\label{second_order}
\ee

We now observe that these two equations can be obtained order by 
order from the expansion around a classical solution, $\phi (t, {\bf x}) 
= \phi_{cl} (t) + \delta \phi^{(1)} + \delta \phi^{(2)}$, of the equation:
\be
\frac{d \phi}{d N} = -\frac{V_\phi}{3 \, H^2} \,,
\label{non_perturb}
\ee
where $N=\log(a/a_0)$ is the e-fold number from the beginning of 
inflation. This suggests that the equation (\ref{non_perturb}) is valid to 
all orders in $\delta \phi \equiv \phi-\phi_{cl}(t)$ in the context of the 
approximation used. Returning to the time evolution and using $dN=H dt$, 
one has:

\ba
\!\!\frac{1}{H}\frac{d}{d t} \delta \phi \!\!&=& -\biggl[
\frac{d}{d \phi} \left(\frac{V_\phi}{3 \, H^2} \right)
\biggl]_{\phi=\phi_{cl}} \!\!\!\!\!\! \delta \phi \nonumber\\
&-&\frac{1}{2} \biggl[
\frac{d^2}{d \phi^2} \left(\frac{V_\phi}{3 \, H^2} \right)
\biggl]_{\phi=\phi_{cl}} \!\!\!\!\!\!\!\!(\delta \phi)^2 +{\cal O}\left( (\delta \phi)^3
\right) \nonumber \\
&\simeq& - \left[ \frac{V_{\phi\phi}}{3 H^2} -
\left(\frac{V_{\phi}}{3 H^2 M_{pl}}\right)^2 \right]_{\phi=\phi_{cl}} 
\!\!\!\!\!\!\!\!\!\! \delta \phi \nonumber \\
 -\frac{1}{2} &{}& \!\!\!\!\!\!\!\!\!\!\!\!
\left[\frac{V_{\phi\phi\phi}}{3H^2}-
\frac{V_{\phi\phi}V_{\phi}}{3 H^4 M_{\rm pl}^2 } + 
\frac{2 V_{\phi}^3}{27 H^6 M_{\rm pl}^4} \right]_{\phi=\phi_{cl}}
\!\!\!\!\!\!\!\!\!\!\!\!\!
(\delta \phi)^2 + {\cal O}\left( (\delta \phi)^3 \right) ,
 \nonumber \\
\ea
where $V_\phi=\frac{dV}{d\phi}$ and so on. The quadratic inflaton
potential leads to the result in Eqs. (\ref{first_order}) and 
(\ref{second_order}) after expanding $\delta \phi=\delta \phi^{(1)}+\delta 
\phi^{(2)}$ and taking the coefficients in the brackets to the 
lowest order in slow-roll. The same method may be used for an arbitrary 
inflaton potential $V(\phi)$.

How does one re-derive these results in the stochastic approach? The 
consideration above suggests that one has to choose the time variable 
$N=\int H(t)dt$ in the Langevin stochastic equation for the large-scale 
part of $\phi$ for this aim. Then it acquires the form (see Eq. (55) of 
\cite{S86}):
\ba
\frac{d \phi}{d N} &=& -\frac{V_\phi}{3H^2} + \frac{f}{H} \,,
\nonumber \\
\langle f(N_1) f(N_2) \rangle &=& \frac{H^4}{4\pi^2}\ \delta(N_1 -N_2) \,, 
\label{phi_general_stoch}
\ea 
where $H^2(t)=V(\phi(t))/3M_{pl}^2$ in this (leading) order of the 
slow-roll approximation, thus, $H$ may be considered as a function of 
$\phi$. As proved in \cite{S86} (see also \cite{SY}), 
though formally the noise $f(N)$ is an operator quantity containing the 
Fock annihilation and creation operators $a_k$ and $a_k^{\dagger}$ for 
modes with $k=\epsilon aH$, its values in different points of space and 
for different $N$ are commutative. Thus, it is equivalent to some 
classical noise whose distribution function appears to be 
Gaussian since the quantum noise $f$ is linear in $a_k$ and $a_k^{\dagger}$. 
Correspondingly, the large-scale part of the quantum inflaton field 
$\hat\phi$ is equivalent to a classical stochastic field $\phi$ with some 
normalized probability distribution $\rho(\phi,N)$, $\int\rho(\phi,N)\, 
d\phi=1$, in the sense of equality of all possible classical and 
quantum expectation values:  
\be
\langle F [ \hat \phi ] \rangle_{\rm REN} 
= \int \rho (\phi)F[\phi] \,d\phi,
\ee   
where $F$ means any functional which may include time and spatial 
derivatives of any order.\footnote{Note that the so called 
volume-weighted averaging \cite{LLM}, for which $\rho$ is not 
normalizable and thus loses the sense of a probability distribution, is 
nowhere used in this paper, as well as in \cite{S86,SY}. For this reason, 
criticism of this approach contained e.g. in the recent paper \cite{P08} 
has no application to our results. See \cite{L07} for the most elaborate 
proposal of how to cure problems of the volume-weighted approach.} 

Following \cite{S86}, the number of e-folds $N$ was considered as a time 
variable in the stochastic Langevin equation in a number of papers, e.g. 
in \cite{Gangui} and most recently in \cite{ENPR08}, while in many other 
ones the proper time $t$ was used, e.g. in \cite{LLM,MM05,GT05,
MartinMusso}. This usage of different time variables should not be mixed 
with the invariance of all physical results with respect to a 
(deterministic) time reparametrization $t\to f(t)$ which is trivially 
satisfied after taking the corresponding change in the metric lapse 
function into account. In contrast, the transformation from $t$ to $N$ made 
using the stochastic function $H(\phi(t))$ leads to a physically 
different stochastic process with another probability distribution.   
Our new statement in this paper following from the arguments above is 
that one {\em should} use the $N$ variable when calculating mean squares
of any quantity containing metric fluctuations like the Mukhanov 
variable $\delta \phi$ or the gauge-invariant metric perturbation
$\zeta$. Otherwise, incorrect results would be obtained using the 
stochastic approach which would then not coincide with those obtained using 
perturbative QFT methods. This statement is further supported by
exact non-perturbative results (valid to all orders of metric
perturbations) from the general $\delta N$ formalism which relates the 
value of $\zeta$ after inflation to the difference in the number of 
e-folds $N$ in different points of space -- it was first used in 
\cite{S82} for one inflaton field, see Eq. (17) of this paper, and then 
generalized to multicomponent inflation in \cite{S85,SS96}, see also
\cite{LR05,LSSTY05} for its recent developments.  
 
On the other hand, the usage of the $t$ variable is natural if one is
interested e.g. in differences of the local duration of inflation in 
proper time in different points of space which, in principle, may be 
measured using a spatial distribution of a phase of the wave function of 
a heavy particle with a mass $m\gg H$. Thus, the choice of a proper time 
variable in the stochastic equation is not an absolute one but is 
dictated by the physical nature of 'clocks' relevant to observable 
effects. In our case, $N$ is the 'clock'. E.g., if a rather small ($\sim 
10\%$) contribution from the integrated Sachs-Wolfe effect due to the 
existence of dark energy (or a cosmological constant) in the present 
Universe is neglected, a large angle anisotropy of the cosmic microwave 
background temperature for multipoles $l<50$ is just given by 
$\Delta N \equiv N - \langle N\rangle$ at the last scattering surface: 
$\Delta T/T = - \Delta N/5$, so that a larger local amount of inflation 
produces a negative spot in the CMB temperature map (here $\langle \rangle$ 
means averaging over the 2-sphere -- the last scattering surface). 

Now, in order to compare with perturbative QFT results, let us assume that
fluctuations in $\phi$ are still much less than the classical background
value $\phi_{cl}(N)$: $|\delta\phi|\equiv |\phi - \phi_{cl}|\ll |\phi_{cl}|$.
In the rest of this section, we shall denote by $H$ and $\phi$ their 
classical background values. Then the noise term in Eq.~
(\ref{phi_general_stoch}) may be considered as a perturbation. In the 
first order, after expanding the first term in the right-hand side of
Eq.~(\ref{phi_general_stoch}) in powers of $\delta\phi$, we get the 
following equation for $\delta\equiv \delta\phi^{(1)}$ (the superscript 
$1$ denotes the order of expansion):
\be
\frac{d\delta}{dN} + 2M_{pl}^2\left(\frac{H'}{H}\right)'\delta
=\frac{f}{H}\,,
\label{delta-eq}
\ee
where the prime denotes the derivative with respect to $\phi$.
Multiplying both sides of Eq.~(\ref{delta-eq}) by $\delta$, averaging and 
using the relation $\langle f\delta\rangle= H^3/8\pi^2$, we obtain the 
equation for $u\equiv \langle(\delta\phi^{(1)})^2\rangle$ corresponding to 
the one-loop approximation of QFT in curved space-time:
\be 
\frac{du}{dN} + 4M_{pl}^2 \left(\frac{H'}{H}\right)' u= 
\frac{H^2}{4\pi^2}\,.
\label{u-eq}
\ee
Eq. (\ref{u-eq}) is valid for any potential $V(\phi)$ satisfying the 
slow-roll conditions. Its generic solution is
\be
u=-\frac{H'^2}{8\pi^2H^2M_{pl}^2}\int \frac{H^5}{H'^3}\,d\phi\,,
\label{u-sol}
\ee 
where the integration variable has been changed from $N$ to $\phi$ using 
the slow-roll relation $dN=-H/(2H'M_{pl}^2)\,d\phi$. Applying this solution 
to the case $V(\phi)=m^2 \phi^2/2,\, H=m\phi/(\sqrt 6 M_{pl})$ and assuming 
$u(0)=0$, we obtain: 
\be
u =  \frac{H_0^{6} - H^{6}}{8 \pi^2 m^2 H^2} \,,
\label{value_phi1_stoch_massive}
\ee
that just coincides with the result derived in \cite{FMVV_II} using QFT 
methods. Also, applying Eq. (\ref{u-sol}) to power-law inflation where
$a(t)\sim t^p$ with $p>>1$ and $H$ exponentially depends on $\phi$, we 
easily recover the correct leading renormalized value of $u$ obtained 
in \cite{Marozzi}.

In the Appendix, the mean value of a second order field fluctuation
$\langle \delta \phi^{(2)} \rangle$ is calculated and also conditions of 
the validity of the perturbation theory are reviewed.

To emphasize the difference, note that would we repeat the same procedure
using the stochastic equation (\ref{phi_general_stoch}) written in terms
of the independent variable $t$, $dt=dN/H$, we would obtain a different
result \cite{LLM,MartinMusso}:
\be
\tilde u = - \frac{H'^2}{8\pi^2M_{pl}^2}\int \frac{H^3}{H'^3}\, d\phi
\label{u-sol-t}
\ee
which, in particular, reduces to $3 (H_0^4-H^4)/(16 \pi^2 m^2)$ for 
massive inflation. As already explained above, the discrepancy between 
(\ref{u-sol}) and (\ref{u-sol-t}) is not surprising and reflects the fact
that different stochastic processes are considered, that is why we
wrote $\tilde u$ instead of $u$ in Eq. (\ref{u-sol-t}).    

The general result presented in Eq. (\ref{u-sol}) for the renormalized value
of the average squared gauge invariant massive inflaton fluctuation has been
explicitely verified by the authors in the context of QFT with renormalization
obtained on employing the adiabatic subtration prescription.
In particular we have considered a family of chaotic inflationary models with
potentials $ V(\phi)=\frac{\lambda}{n} \kappa^{n-4} \phi^n$ and obtained for
such models results similar to those presented in section II, but valid for
gauge invariant fluctuations. Eq. (\ref{u-sol}) has been found to give
renormalized values in complete agrement with our perturbative QFT
calculations in the slow roll approximation.

\section{4-loop calculation for a massless self-interacting test
scalar field in the stochastic approach}
In the previous sections, only the mean value of 
$\langle(\delta\phi^{(1)})^2\rangle$ corresponding to one scalar loop
in external background curved space-time was calculated (with small metric 
fluctuations taken into account also for the case of an inflaton scalar 
field). However, as was shown in \cite{S86,SY}, the stochastic approach can 
reproduce QFT results for any finite number of scalar loops and even 
beyond (e.g. results obtained using instantons). As an example, let us 
consider a massless self-interacting test scalar field $\chi$ with the
potential $V(\chi)=\lambda \chi^4/4$ in the exact de Sitter space-time 
with the curvature $H_0$ following \cite{SY}. Then $N=H_0t$, and it makes
no difference which time variable is used in the stochastic Langevin 
equation (\ref{phi_general_stoch}). It is straightforward to construct 
the corresponding Fokker-Planck equation for a normalized probability 
distribution $\rho(\chi,N)$ \footnote{This one point distribution function 
should not be confused with the energy density $\rho_\chi$ of a test 
$\chi$ field considered in Sects. II-V.}:
\begin{equation}
\frac{\partial\rho}{\partial N} = \frac{\lambda}{3 H_0^2} \frac{\partial
\left( \rho \chi^3 \right)}{\partial \chi} + \frac{H_0^2}{8 \pi^2}
\frac{\partial^2 \rho}{\partial \chi^2}\,.
\label{FP-eq}
\end{equation}

Multiplying Eq. (\ref{FP-eq}) by $\chi^n$ where $n$ is an even integer
and integrating over $\chi$ from $-\infty$ to $\infty$, we obtain the
following recurrence relation (c.f. also \cite{TW05}): 
\be 
\frac{d}{dN} \langle \chi^n\rangle = - \frac{n\lambda}{3H_0^2}
\langle\chi^{n+2}\rangle + \frac{n(n-1)H_0^2}{8\pi^2}
\langle\chi^{n-2}\rangle \,.
\label{recur}
\ee
Then solving Eq. (\ref{recur}) iteratively beginning from $n=2$ with the 
initial conditions $\langle\chi^n(0)\rangle =0$ for all $n$, we find:
\begin{eqnarray}
\langle \chi^2 \rangle_{\rm REN} &=& \frac{H_0^2 N}{4
\pi^2} \left( 1 + \alpha_1 \lambda X^2 + \alpha_2 \lambda^2 X^4 +
\alpha_3 \lambda^3 X^6 + ... \right) \,,
\nonumber \\
\langle \chi^4 \rangle_{\rm REN}
&=& 3 \left(\frac{H_0^2 N}{4
\pi^2} \right)^2 \left( 1 + \beta_1 \lambda X^2 + \beta_2 \lambda^2 X^4
+ ... \right)\,, 
\nonumber\\
\langle \chi^6 \rangle_{\rm REN}
&=& 15 \left(\frac{H_0^2 N}{4
\pi^2} \right)^3 \left( 1 + \gamma_1 \lambda X^2 + ... \right)\,, 
\label{self_desitter}
\end{eqnarray}
where $X=N/(2 \pi)$
and
\begin{eqnarray}
\alpha_1 &=& - \frac{2}{3} \quad \quad \alpha_2 = \frac{4}{5} \quad \quad
\alpha_3 = - \frac{424}{315} \nonumber \\
\beta_1 &=& - 2 \quad \quad \beta_2 = \frac{212}{45} \quad\quad \gamma_1=-4\,.
\label{coeff}
\end{eqnarray}
The result for $\alpha_1$ agrees with the perturbative QFT computations in 
\cite{OW,BOW} (see also \cite{KO}). Since the $\lambda$-independent term
in the expression for $\langle \chi^2 \rangle_{\rm REN}$ corresponds to one
scalar loop and each next power of $\lambda$ requires one more scalar loop,
the total result for $\langle \chi^2 \rangle_{\rm REN}$ presented in
(\ref{self_desitter}) and (\ref{coeff}) requires consideration of 
diagrams up to 4 loops included in the perturbative QFT treatment.

\section{Conclusions}
We have studied the application of the stochastic approach to 
inflationary space-times with $\dot H \ne 0$ and compared its results 
to those obtained using the standard QFT in curved space-time,
extending the results of \cite{FMVV_I,FMVV_II}. We can summarize our 
main results by the following points:

1. On considering a test field $\chi$ with mass $m_\chi$ in $m^2 \phi^2$ 
inflation, we have shown that the instantaneous Bunch-Davies expectation 
value $\langle \chi^2 \rangle = 3 H^4(t)/(8 \pi^2 m^2_\chi)$ is never 
reached. It may be approached at the end of inflation for $m_\chi >> m$ 
only. If $m_\chi << m$, the value of $\langle \chi^2 \rangle$ at the end 
of inflation is quite different from that extrapolated from the
exact de Sitter space-time. 

2. We have analysed implications of the particle production peculiar 
to $m^2 \phi^2$ inflation. The moduli problem is more serious 
than in the classical counterpart and also with respect to previous 
quantum investigations \cite{FKL}.  

3. Concerning gauge-invariant inflaton fluctuations, we have clarified 
why the stochastic Langevin equation for the large-scale part of an 
inflaton field should be formulated using the number of e-folds 
$N=\ln (a/a_0)$ as an independent time variable, if one is interested in 
any result regarding gauge-invariant inflaton fluctuations and metric
perturbations. The mean square of inflaton fluctuations calculated in 
this way has been shown to coincide with the earlier result of 
\cite{FMVV_II} obtained using standard perturbative methods.

4. In the inflationary space-times studied here, neither test fields nor 
inflaton mean square expectation value admit a static equilibrium solution.

5. The equivalence between the stochastic and the standard field-theoretic
approaches works beyond the one-loop approximation to QFT in curved 
space-time. 

\vspace{1cm}

{\bf Acknowledgements}

AS was partially supported by the Russian Foundation for Basic Research,
grant 08-02-00923, and by the Research Programme "Elementary Particles"
of the Russian Academy of Sciences. This work was started during visits 
of AS to Bologna in 2003-5 financed by INFN: 
we thank INFN for support. Some of these results were presented 
at the International Conference on Theoretical Physics in the Physical 
Lebedev Institute, Moscow (Russia), April 11-16, 2005.

\section{Appendix}
In this appendix, we shall denote by $H$ and $\phi$ their
classical background values; we also use 
$u\equiv \langle(\delta\phi^{(1)})^2\rangle$.
Starting from Eq.(\ref{phi_general_stoch}), and performing a
perturbative expansion,
it is easy to derive for the average second order fluctuation 
$\langle \delta \phi^{(2)} \rangle$ a time evolution also
governed by (see also \cite{Gangui})
\begin{widetext}
\be
\frac{d}{d t} \langle \delta \phi^{(2)} \rangle
=\frac{H^3}{16 \pi^2} \left(\frac{V_\phi}{V}\right)
-\left(\frac{1}{3H} V_{\phi\phi} +2 \frac{\dot{H}}{H} 
\right)
\langle \delta \phi^{(2)} \rangle
+\frac{1}{2}\left[ -\frac{1}{3H} V_{\phi\phi\phi}+
\left( \frac{1}{H} V_{\phi\phi} +4
  \frac{\dot{H}}{H}\right) \frac{V_\phi}{V}
\right]\langle u
\label{phi2_evol_stoch}
\ee
The general solution of Eq. (\ref{phi2_evol_stoch}) with the 
initial condition $\delta \phi (t=t_i)=0$ 
is given by
\be
\langle \delta \phi^{(2)} \rangle =\left(\frac{V_\phi}{V}
\right) \int_{t_i}^t dt' \left(\frac{V}{V_\phi}
\right)\left\{ \frac{H^3}{16 \pi^2} 
\left(\frac{V_\phi}{V}
\right)+\frac{1}{2}
\left[ -\frac{1}{3H} V_{\phi\phi\phi}+
\left( \frac{1}{H} V_{\phi\phi} +4
  \frac{\dot{H}}{H}\right) \frac{V_\phi}{V}
\right]\langle u
\rangle\right\}\,.
\label{phi2_sol_stoch}
\ee

\end{widetext}

In the particular case of a chaotic inflation with 
$V(\phi)=\frac{m^2}{2}\phi^2$ we obtain Eq. (\ref{value_phi1_stoch_massive}) 
for first order and: 
\be
\langle \delta \phi^{(2)} \rangle =  
\frac{\dot{\phi}}{32 \pi^2 m^2 H M_{pl}^2}\left[
\frac{H_0^{6} -H^{6}}{H^2} -3 \left(H_0^{4} -H^{4} \right)
\right]\,,
\label{value_phi2_stoch_massive}
\ee
for second order. Note that 
the solution (\ref{value_phi1_stoch_massive}) 
clearly agrees with the solution (\ref{soldiff}) 
for suitable initial conditions and $\alpha=-1$.

Now we wish study the validity of the perturbative expansion by considering
the ratio $\frac{\langle \delta \phi^{(2)} \rangle}{\sqrt{u}}$
and $\frac{\sqrt{u}}{\phi}$.
Using the former results one obtains
\be
\frac{\langle \delta \phi^{(2)} \rangle}{\sqrt{u}}
=-\frac{1}{8 \pi \sqrt{3}}
\frac{1}{M_{pl}}\frac{\frac{1}{H^2}
\left(H_0^{6} -\!H^{6} \right)\!-\!3 \left(H_0^{4} -\!H^{4} \right)}{\left(
H_0^{6} -H^{6} \right)^{1/2}}
\label{ratio2_1}
\ee
\be
\frac{\sqrt{u}}{\phi}=\frac{1}{4 \pi \sqrt{3}}
\frac{1}{M_{pl}}\frac{1}{H^2}\left(
H_0^{6} -H^{6} \right)^{1/2}\,.
\label{ratio1_0}
\ee
Thus, as we can see, those two ratios differ only by a factor $2$ 
for the leading term
toward the end of inflation.
To see when the perturbative expansion is no longer useful we can use the 
variable $\tilde{N}$ defined as the number of e-folds away from the 
maximum value 
$N_{max}=N_0=\log{\frac{a_{max}}{a(t_i)}}=\frac{3}{2}\frac{H_0^{2}}{m^2}$ 
namely
\be
N_{max}-\tilde{N}=\log{\frac{a(t)}{a(t_i)}} \rightarrow 
\tilde{N}=\frac{3}{2}\frac{H^{2}}{m^2}\,.
\label{Ntilde}
\ee
Using this variable one obtains, to leading order, for the ratio 
(\ref{ratio2_1}) the 
following result
\be
\frac{\langle \delta \phi^{(2)} \rangle}{\sqrt{u}}=-\frac{\sqrt{2}}{24 \pi}
\frac{m}{M_{pl}}\frac{N_{max}}{\tilde{N}} \left(N_{max}-\tilde{N}
\right)^{1/2}\,.
\label{ratio2_1_Ntilde}
\ee
If we require that the absolute value of this ratio be less then one we obtain,
under the condition $768 \pi^2 \frac{M_{pl}^2}{H_0^2}>>1$,
the following approximate constraint

\be
\tilde{N} \ge \frac{3 \sqrt{3}}{48 \pi}\frac{H_0^3}{M_{pl} m^2}\,.
\label{solution2_1_Ntilde}
\ee
If we consider the end of inflation when $H=m$ this corresponds to 
$\tilde{N}=3/2$, so in order to have a perturbative expansion 
with small terms for all of the 
duration of inflation we obtain, on considering the condition 
(\ref{solution2_1_Ntilde}) with $\tilde{N}=3/2$, the following 
condition on $H_0$:
\be 
H_0<\left(\frac{24 \pi}{\sqrt{3}}M_{pl} m^2
\right)^{1/3}\,,
\label{condition_H0_2_1}
\ee
similarly on considering the ratio (\ref{ratio1_0}) one obtains the 
condition 
\be 
H_0<\left(\frac{12 \pi}{\sqrt{3}}M_{pl} m^2
\right)^{1/3}\,,
\label{condition_H0_1_0}
\ee
the above correspond, for the particular numerical value $M_{pl}=10^5 m$, 
respectively to $H_0<163.28 m$ and 
$H_0<129.596 m$. Such bounds are in agreement with previous investigations
\cite{FMVV_IV}.



\begin{thebibliography}{10}

\bibitem{bd_book}
N.~D.~Birrell and P.~C.~W. Davies, {\em Quantum Fields in Curved Space}
(Cambridge University Press, Cambridge, 1982).

\bibitem{L82}
A.~D.~Linde, Phys.\  Lett.\  B {\bf 116}, 335 (1982).

\bibitem{S82}
A.~A.~Starobinsky, Phys.\ Lett.\ B {\bf 117}, 175 (1982).

\bibitem{VF82}
A.~Vilenkin and L.~Ford, Phys.\ Rev.\ D {\bf 26}, 1231 (1982).

\bibitem{GH77}
G.~W.~Gibbons and S.~W.~Hawking, Phys.\ Rev.\ D {\bf 15}, 2738 (1977).

\bibitem{S86} 
A.~A.~Starobinsky, in {\em Field Theory, Quantum Gravity
and Strings}, eds. H.~J.~De~Vega and N.~Sanchez, Lect.\ Notes\
in\ Physics {\bf 246}, 107 (1986).

\bibitem{V83}
A.~Vilenkin, Phys.\ Rev.\ D {\bf 27}, 2848 (1983).

\bibitem{L86}
A.~D.~Linde, Phys.\ Lett.\ B {\bf 175}, 395 (1986).

\bibitem{SY}
A.~A.~Starobinsky and J.~Yokoyama, Phys.\ Rev.\ D {\bf 50}, 6357 (1994).

\bibitem{FMVV_I}
F.~Finelli, G.~Marozzi, G.~P.~Vacca and G.~Venturi,
Phys. Rev. {\bf D 65}, 103521 (2002).

\bibitem{FMVV_II}
F.~Finelli, G.~Marozzi, G.~P.~Vacca, and G.~Venturi,
Phys.\ Rev.\ D {\bf 69}, 123508 (2004).

\bibitem{GLV}
A.~S.~Goncharov, A.~D.~Linde and M.~I.~Vysotsky, Phys.\ Lett.\  B 
{\bf 147} 279 (1984).

\bibitem{FKL}
G.~N.~Felder, L.~Kofman and A.~D.~Linde, JHEP {\bf 0002}, 027 (2000). 

\bibitem{S78}
A.~A.~Starobinsky, Sov.\ Astron.\ Lett. {\bf 4}, 82 (1978).

\bibitem{KLS85}
L.~A.~Kofman, A.~D.~Linde and A.~A.~Starobinsky, Phys.\ Lett.\ B
{\bf 157}, 361 (1985).

\bibitem{GMS91}
S.~Gottl\"ober, V.~M\"uller and A.~A.~Starobinsky, Phys.\ Rev.\ D 
{\bf 43}, 2510 (1991). 

\bibitem{FVV}
F.~Finelli, G.~P.~Vacca and G.~Venturi, Phys.\ Rev.\  D {\bf 58}, 
103514 (1998).

\bibitem{LM}
A.~D.~Linde and V.~F.~Mukhanov, Phys.\ Rev.\  D {\bf 56}, 535 (1997).

\bibitem{MT01}
T.~Moroi and T.~Takahashi, Phys.\ Lett.\  B {\bf 522}, 215 (2001).

\bibitem{ES02}
K.~Enqvist and M.~S.~Sloth, Nucl.\ Phys.\ B {\bf 626}, 395 (2002).

\bibitem{LW}
D.~H.~Lyth and D.~Wands, Phys.\ Lett.\  B {\bf 524}, 5 (2002).

\bibitem{KT}
E.~W.~Kolb and M.~S.~Turner, {\em The Early Universe} (Addison-Wesley,
Redwood City, California, 1990).

\bibitem{M85}
V.~F.~Mukhanov, JETP\ Lett. {\bf 41}, 493 (1985).

\bibitem{FMVV_IV}
F.~Finelli, G.~Marozzi, G.~P.~Vacca and G.~Venturi, Phys.\ Rev.\ D 
{\bf 74}, 083522 (2006).

\bibitem{LLM}
A.~D.~Linde, D.~A.~Linde and A.~Mezhlumian, Phys.\ Rev.\  D {\bf 49}, 
1783 (1994).

\bibitem{P08}
D.~N.~Page, JCAP {\bf 0810}, 025 (2008).

\bibitem{L07}
A.~D.~Linde, JCAP {\bf 0706}, 017 (2007).

\bibitem{Gangui}
A.~Gangui, F.~Lucchin, S.~Matarrese and S.~Mollerach, Astroph.\ J. 
{\bf 430}, 447 (1994).

\bibitem{ENPR08}
K.~Enqvist, S.~Nurmi, D.~Podolsky and G.~I.~Rigopoulos,
JCAP {\bf 0804}, 025 (2008).

\bibitem{MM05}
J.~Martin and M.~Musso, Phys.\ Rev.\ D {\bf 71}, 063514 (2005).

\bibitem{GT05}
S.~Gratton and N.~Turok, Phys.\ Rev.\ D {\bf 72}, 043507 (2005).

\bibitem{MartinMusso}
J.~Martin and M.~Musso, Phys.\ Rev.\ D {\bf 73}, 043516 (2006).

\bibitem{S85}
A.~A.~Starobinsky, JETP\ Lett. {\bf 42}, 152 (1985).

\bibitem{SS96}
M.~Sasaki and E.~D.~Stewart, Progr.\ Theor.\ Phys. {\bf 95}, 71 (1996).

\bibitem{LR05}
D.~H.~Lyth and Y.~Rodriguez, Phys.\ Rev.\ Lett. {\bf 95}, 121302 (2005).

\bibitem{LSSTY05}
H.-C.~Lee, M.~Sasaki, E.~D.~Stewart, T.~Tanaka and S.~Yokoyama,
JCAP {\bf 0510}, 004 (2005).

\bibitem{Marozzi}
G.~Marozzi, Phys.\ Rev.\ D {\bf 76}, 043504 (2007).

\bibitem{TW05}
N.~G.~Tsamis and R.~P.~Woodard, Nucl.\ Phys.\ B {\bf 724}, 295 (2005).

\bibitem{OW}
V.~K.~Onemli and R.~P.~Woodard, Class.\ Quant.\ Grav. {\bf 19}, 
4607 (2002).

\bibitem{BOW}
T.~Brunier, V.~K.~Onemli and R.~P.~Woodard, Class.\ Quant.\ Grav. 
{\bf 22}, 59 (2005).

\bibitem{KO}
E.~O.~Kahya and V.~K.~Onemli, Phys.\ Rev.\ D {\bf 76}, 043512 (2007). 

\end{thebibliography}
\end{document}